\documentclass[iop,apj]{emulateapj}

\usepackage{soul,color}
\usepackage{amssymb,amsmath}

\def\lea{\mathrel{<\kern-1.0em\lower0.9ex\hbox{$\sim$}}}
\def\gea{\mathrel{>\kern-1.0em\lower0.9ex\hbox{$\sim$}}}
\newcommand{\lta}{{\>\rlap{\raise2pt\hbox{$<$}}\lower3pt\hbox{$\sim$}\>}}
\newcommand{\gta}{{\>\rlap{\raise2pt\hbox{$>$}}\lower3pt\hbox{$\sim$}\>}}

\begin{document}
\title{X-ray Binaries and Star Clusters in the Antennae: Optical Cluster Counterparts}

\author{Blagoy Rangelov\altaffilmark{1}, Rupali Chandar\altaffilmark{1}, Andrea Prestwich\altaffilmark{2}, and Bradley C. Whitmore\altaffilmark{3}} 
\altaffiltext{1}{Department of Physics and Astronomy, The University of Toledo, Toledo, OH 43606, blagoy.rangelov@gmail.com}
\altaffiltext{2}{Harvard-Smithsonian Center for Astrophysics, 60 Garden Street, Cambridge, MA 02138}
\altaffiltext{3}{Space Telescope Science Institute, 3700 San Martin Drive, Baltimore, MD 21218}

\slugcomment{The Astrophysical Journal, in press}
\shorttitle{XRBs and star clusters in the Antennae}
\shortauthors{Rangelov et al.}

\begin{abstract}

We compare the locations of 82 X-ray binaries (XRBs) detected in the merging Antennae galaxies by Zezas et al.,based on observations taken with the {\em Chandra X-Ray Observatory}, with a catalog of optically selected star clusters presented by Whitmore et al., based on observations taken with the {\em Hubble Space Telescope}. Within the $2\sigma$ positional uncertainty of $\approx0\farcs58$, we find 22 XRBs are coincident with star clusters, where only $2-3$ chance coincidences are expected. The ages of the clusters were estimated by comparing their $UBVI$, H$\alpha$ colors with predictions from stellar evolutionary models. We find that 14 of the 22 coincident XRBs (64\%) are hosted by star clusters with ages of $\approx 6$~Myr or less. Five of the XRBs are hosted by young clusters with ages $\tau \approx 10-100$~Myr, while three are hosted by intermediate age clusters with $\tau \approx 100-300$~Myr. Based on the results from recent $N$-body simulations, which suggest that black holes are far more likely to be retained within their parent clusters than neutron stars, we suggest that our sample consists primarily of black hole binaries with different ages.

\end{abstract}


\section{Introduction}

Observations of the ``Antennae'' (NGC~4038/39) reveal that this merging system of two gas-rich spiral galaxies is full of point-like X-ray sources, many of which are probably high-mass X-ray-emitting binary star systems (e.g., \citet{Zezas2002}). Most X-ray binaries (hereafter XRBs) in star-forming galaxies are believed to have either a black hole (BH) or a neutron star (NS) as the compact source, formed after a massive star ends its life as a supernova. If this remnant is in a binary system with another star, mass transfer onto the compact object can result in X-ray emission. In this paper, we will refer to XRBs that have a massive donor star as high-mass X-ray binaries (HMXBs).

The Antennae have a fairly high rate of star formation, $\approx20-30~M_{\sun}$~yr$^{-1}$, with much of this star formation occurring in compact star clusters (e.g., \citet{Whitmore_chweizer1995,Whitmore1999,Fall2005,Whitmore2010}). \citet{Fall2005} found that \emph{at least  20\%}, and possibly 100\% of the massive stars in the Antennae form in compact star clusters. Most O stars in the Galaxy are also found in clusters and associations. Studies of Galactic O stars currently found in the {\em field} suggest that all but 4\% of these likely formed in nearby clusters and associations, but were subsequently ejected \citep{deWit2004,deWit2005}. Given the strong constraints on the (high) fraction of massive stars formed in clusters, we consider it highly likely that HMXBs in the Antennae formed within star clusters.

Once formed, binary systems can: (1)  remain within their host cluster, (2) be dynamically kicked out of their parent/host cluster (either due to the supernova explosion or due to dynamical interactions with other stars in the crowded centers of star clusters), or (3) be left behind after the parent cluster dissolves. The first scenario leads to the strongest constraint on the ages of XRBs, via the age of the parent/host star cluster. Therefore, precise age estimates for host clusters can provide important information on XRBs. This paper focuses on XRBs that remain within their host cluster in the Antennae. The last two scenarios, which lead to a situation where XRBs are close to but not coincident with star clusters, will be studied in a second paper (Paper~II).

\citet{Clark2011} recently compared X-ray sources in the Antennae compiled by \citet{Zezas2006} with near-infrared images from the WIRC camera on the Palomar 5-m telescope, and optical images from the WFPC2 camera on the \emph{Hubble Space Telescope} (\emph{HST}). The FWHM of the IR images is approximately $1\arcsec$, and that of the optical images is $\approx0\farcs522$ (the three WF CCDs on the WFPC2 camera have a plate scale of $0\farcs51$~pix$^{-1}$). \citet{Clark2011} found 32 likely IR counterparts to the X-ray sources. They estimated ages for a subset of 10 sources by comparing integrated $UBVIJK$ photometry with predictions from the STARBURST99 stellar population models \citep{Leitherer1999}, and derived ages for all 10 clusters of log$(\tau/\mbox{yr}) \approx6.9-7.5$. Their derived masses are $M \gea 10^5~M_{\odot}$ with several clusters having estimated masses of $\approx 10^6~M_{\odot}$.

\citet{Whitmore2010} recently studied star clusters in the Antennae from deeper, higher resolution observations taken with the Advanced Camera for Surveys (ACS) camera on $HST$ (the data and observations are summarized in Section~2.2). The availability of this new cluster catalog allows us to revisit the properties of star clusters in the Antennae that host XRBs, using the highest quality observations currently available. In this paper, we investigate the ages and masses of star clusters that are coincident with X-ray sources in the merging Antennae galaxies. We assume a distance of 21 Mpc to the Antennae (distance modulus of 31.71 mag; \citet{Schweizer2008,Riess2011}). At this distance, $1\arcsec$ subtends a distance of 99.8 pc. Just over a quarter of the likely HMXBs in the Antennae are coincident or nearly coincident with a star cluster, and are the focus of this work. A follow-up paper will investigate the relationship between star clusters and XRBs for the remaining three-quarters of the population that are not coincident with a star cluster.

This paper is organized as follows. Section~2 summarizes the X-ray observations and analysis of the XRBs performed by \citet{Zezas2006}. It also summarizes the optical observations and analysis of the star clusters performed by \citet{Whitmore2010}. Section~3 describes the astrometric matching  between the X-ray and optical source catalogs. Section~4 presents the properties of star clusters that are coincident with XRBs in the Antennae, and Section~5 discusses the implications of these results for the nature of the XRBs. We summarize our main conclusions in Section~6.

\section{Data and Source Catalogs}

\subsection{X-Ray Observations and Catalog of XRBs}

The Antennae were observed with the ACIS-S instrument on the \emph{Chandra X-Ray Observatory} as part of two programs (PI: Murray, Proposal Number: 01600062; and PI: Fabbiano, Proposal Number: 03700413). There are seven individual observations with integration times between 37 and 75 ks, and a total exposure time of 411 ks. The limiting luminosity of the observations range between $\thicksim2\times10^{37}$ and $5\times10^{37}$ erg s$^{-1}$, depending on the exposure time and local background (see \citet{Zezas2006} for more details). The publicly available online catalog \citep{Zezas2006} contains seven more sources that were detected in a merged exposure, thus increasing the total number to 127. Of these, 96 are found within the main body and nuclear region of the Antennae and hence are likely to be associated with the merger, while the rest are almost certainly background galaxies.

The \emph{Chandra} observations of the Antennae galaxies reveal both point sources and diffuse X-ray emission. \citet{Zezas2006} suggested that 15 of their sources may be detections of diffuse emission rather than point sources. We visually inspected these sources in the observations, and agree that these detections are unlikely to be XRBs except in two cases (X84 and X94), which have counts in the hard band and are therefore unlikely to be associated with the diffuse plasma. The remaining 13 sources are eliminated from further consideration. We also eliminate X28,  because we could not distinguish it spatially from X27 in the \emph{Chandra} observations. This leaves a sample of 82 candidate XRBs in the Antennae. We show the locations of these sources on an optical image of the Antennae in Figure 1.

%%%%%%%%%%%%%%%%%%%%
\begin{figure*}
\epsscale{1}
\plotone{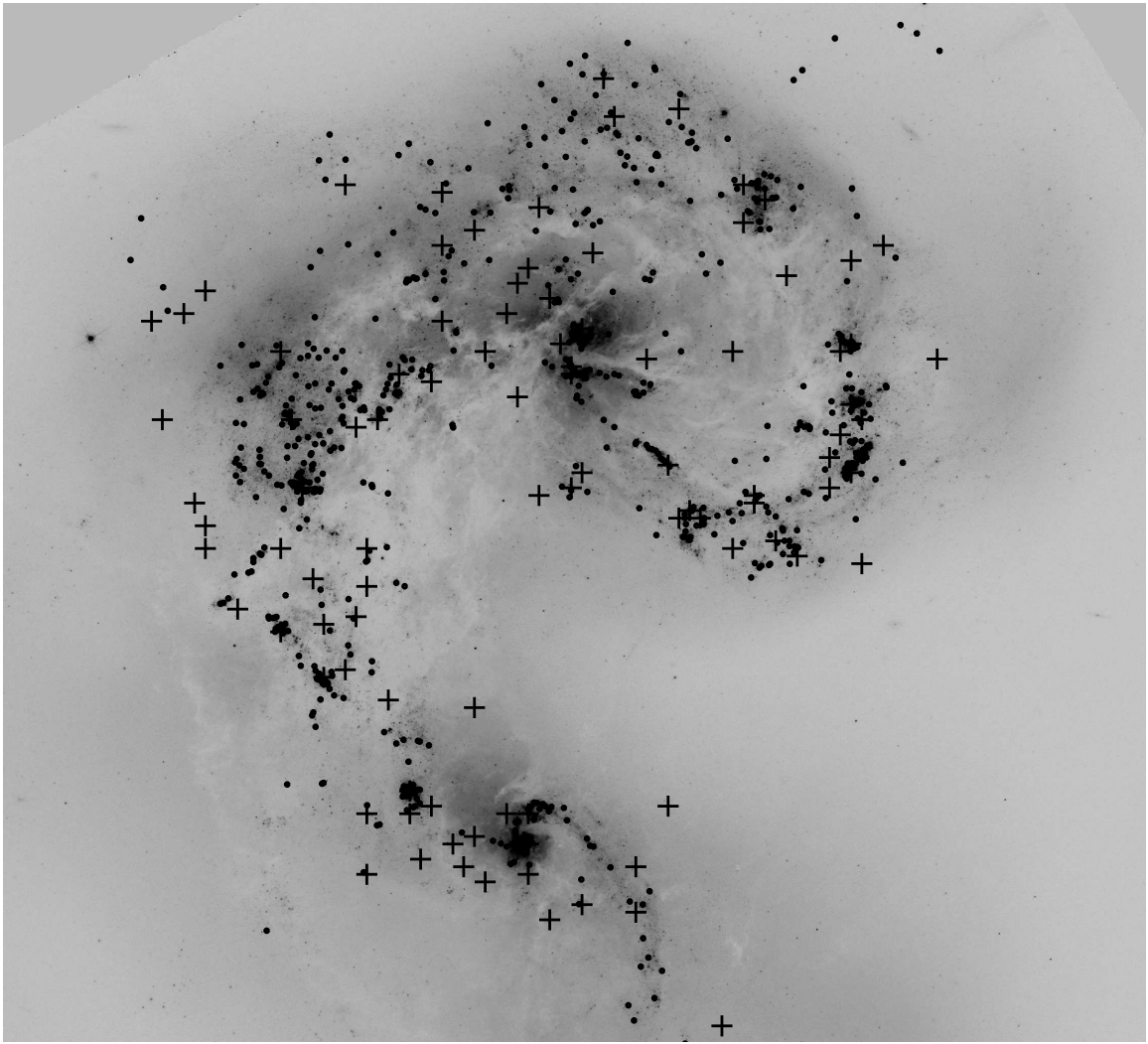}
\caption{
Optical image of the Antennae galaxies taken with the ACS camera on the $HST$. North is up and east is to the left. The locations of the star clusters used in this work are shown as circles. The locations of X-ray sources selected from $Chandra$ X-ray observations are shown as crosses. See the text for details.
\label{fig-1}}
\end{figure*}
%%%%%%%%%%%%%%%%%%%%

%%%%%%%%%%%%%%%%%%%%
\begin{figure*}
\epsscale{1}
\plotone{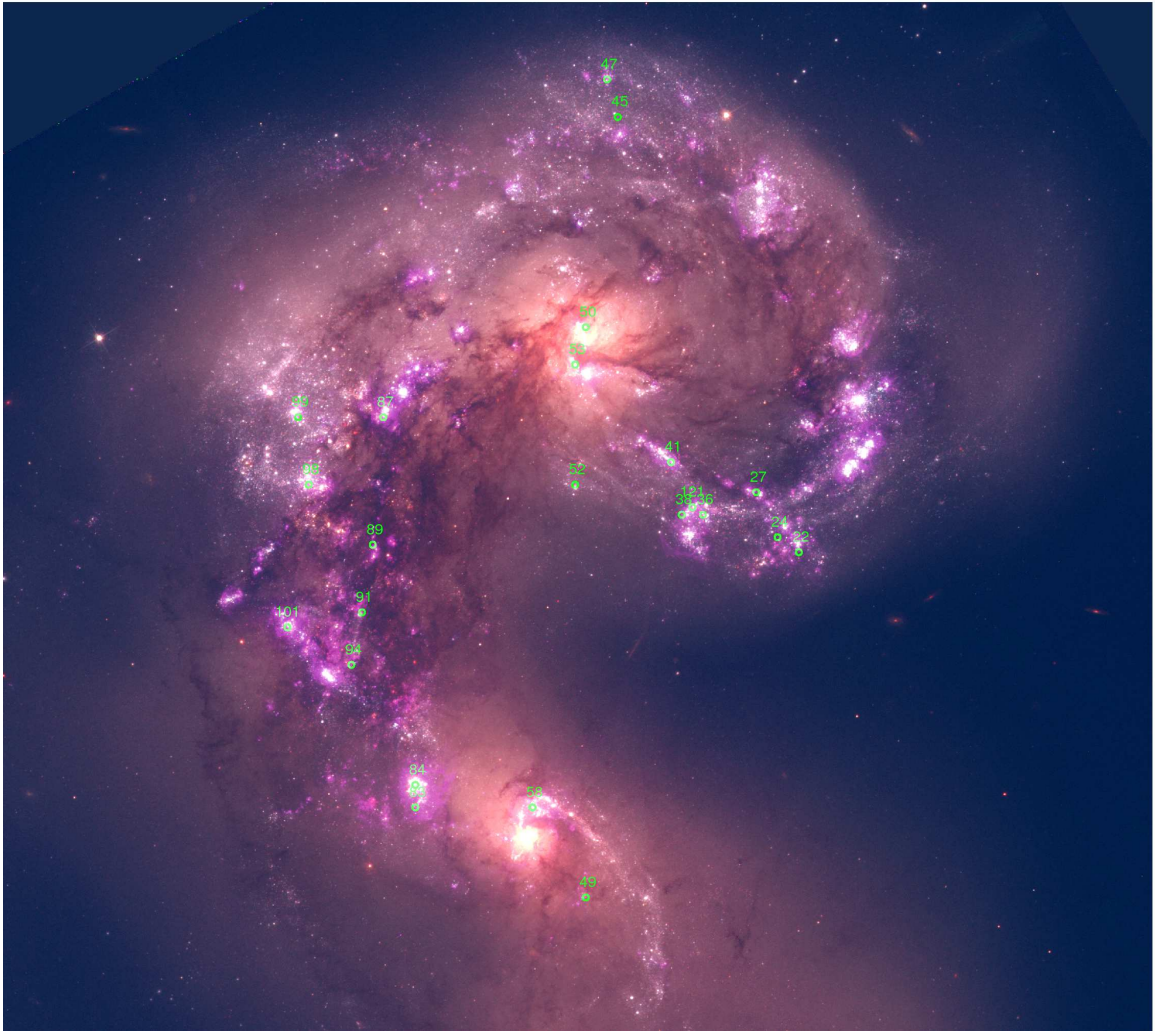}
\caption{Color image of the Antennae produced using \emph{HST}/ACS images in $B,V,I$, and the narrow-band H$\alpha$ observations (shown in purple) to highlight the sites of recent cluster formation. The locations of the 22 XRBs that are coincident with a star cluster, and which are the focus of this paper, are identified and labeled.
\label{fig-2}}
\end{figure*}
%%%%%%%%%%%%%%%%%%%%

%%%%%%%%%%%%%%%%%%%%
\begin{figure*}
\epsscale{1}
\plotone{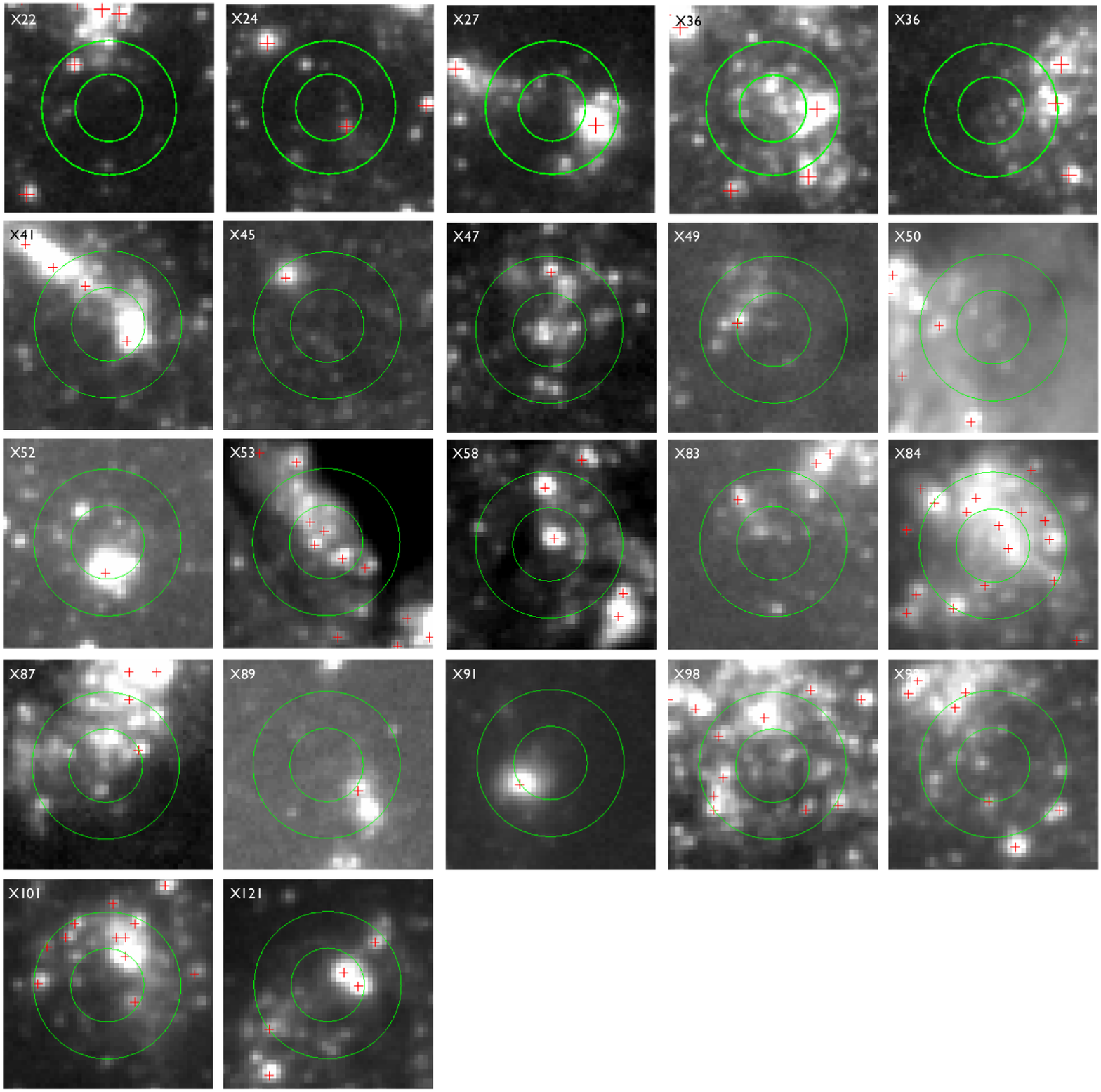}
\caption{
A $2\farcs55\times2\farcs55$ region around each of the 22 coincident sources is shown from the \emph{HST} $V$-band image. The two (green) circles represent  $0\farcs54$ and $0\farcs58$ ($1\sigma$ and $2\sigma$) positional uncertainties around each XRB. The (red) crosses identify clusters within the vicinity of each XRB.
%Each image is $2\arcsec\times2\arcsec$.
\label{fig-3}}
\end{figure*}
%%%%%%%%%%%%%%%%%%%%

\citet{Zezas2006} extracted a number of parameters for the X-ray sources in the Antennae, including locations, counts in the ``soft'' ($S$, $0.3-1.0$~keV), ``medium'' ($M$, $1.0-2.5$~keV), and ``hard'' ($H$, $2.5-7.0$~keV) bands in each individual exposure, and from the combined image. Following \citet{Prestwich2003}, we define two X-ray colors, $H1$ and $H2$, as follows. The soft color is defined as $H1 = (M - S)/T$, and the hard color as $H2 = (H - M)/T$, where $S$, $M$ and $H$ are the total counts in the bands defined above, and $T$ is the total number of counts in all three bands. The $H1$ and $H2$ values are determined from the published values in \citet{Zezas2006}, after subtracting the corresponding background counts for each band. Note that the energy ranges in the $S$, $M$, and $H$ bands are defined somewhat differently from those in Prestwich et al. (2003).

\subsection{Optical Observations and Catalog of Star Clusters}

The XRBs detected in the Antennae will be compared with optically\footnote{Optically selected cluster catalogs directly detect approximately 85\% of clusters younger than a few Myr, i.e. the Antennae do not appear to contain a significant population of very young, highly obscured star clusters \citep{Whitmore_Zhang2002}.} selected star clusters presented in \citet{Whitmore2010}. This cluster catalog was based on $HST$ observations of the Antennae taken with the ACS/WFC camera in the broad-band $BVI$ and narrow-band H$\alpha$ filters. The WFC on ACS has a pixel scale of $0\farcs505$~pixel$^{-1}$. Point-like sources were detected using the DAOFIND task in IRAF, and photometry was performed using the IRAF task PHOT, with a 2 pixel aperture radius and inner and outer background annuli of 4 and 7 pixels, respectively.  Corrections were applied to convert the aperture magnitudes to total magnitudes. Here, the H$\alpha$ filter contains both stellar continuum and nebular line emission (i.e., we did not perform any continuum subtraction). The instrumental magnitudes were converted to the VEGAMAG system using zeropoints from \citet{Sirianni2005} for ACS. In addition, $U$-band photometry was measured for most sources from $HST$/WFPC2 images (the three WFC CCDs have a pixel scale of 0\farcs51 and the PC CCD has 0.05''), using a 2 pixel aperture for the WFC CCDs (1.5~pixels for the PC) with zeropoints from \citet{Holtzman1995}. The full catalog contains both individual stars and star clusters in the Antennae. Here, we select star clusters by restricting the sample to sources brighter than $M_V$ of $-9$, a limit that is brighter than nearly all individual stars.

We estimated the age and extinction of each cluster by performing a least $\chi^2$ fit comparing their $UBVI$,H$\alpha$ magnitudes with predictions from the \citet{BC2003} models for simple stellar populations, assuming solar metallicity, a Salpeter initial stellar mass function, and a Galactic-type extinction curve. The mass of each cluster was estimated from the extinction-corrected $V$ band luminosity and the (age-dependent) mass-to-light ratio predicted by the Bruzual \& Charlot models, assuming a distance modulus of $31.71$~mag to the Antennae \citep{Schweizer2008,Riess2011}.

In \citet{Whitmore2010} we assessed the accuracy of our photometric age estimates by comparing with ages determined for 16 clusters from ground-based spectra \citep{Bastian2009}. We found that our age estimates are generally within log~$(\tau/ \mbox{yr}) \approx0.3$ or a factor of two of the spectroscopic ones for the age range of relevance here, $\tau \lea 3\times 10^8$~yr.

For this work, we also created a continuum-subtracted H$\alpha$ image of the Antennae, which reveals the sites of recent star formation. An examination of this image shows that clusters with estimated ages of $\tau \lea 6$~Myr are  associated with nebular emission in nearly all cases, as expected, again supporting the accuracy of our cluster age determinations. This image will be used in Section~4 to investigate the presence or absence of ionized gas associated with star clusters that are coincident with XRBs. For the mass estimates, the largest uncertainty comes from uncertainty in the age determinations, which dominate the predicted mass-to-light ratio. We previously estimated that the cluster masses are accurate to within approximately a factor of two.

\section{Astrometric Matching of the X-ray and Optical Catalogs}

In order to find optical counterparts to XRBs in the Antennae, we first need to match the X-ray and optical coordinate systems. We use the R.A. and decl. of X-ray sources presented in \citet{Zezas2006}. For the optical data, we use the combined $BVI$ image available from the Hubble Legacy Archive (HLA,l http:\/www.hla.stsci.edu); sources in the HLA images have excellent {\em relative} positions, but there may be global shifts, at the $\approx 1-2\arcsec$ level,
relative to their absolute positions.

We first identified the optical counterparts to 10 X-ray sources (X1, X2, X3, 6, X8, X37, X60, X90, X107, and X117) in the outer regions of the HLA image. None of these sources are located within the main body of the Antennae, and hence are almost certainly background or foreground sources. Source X90 is a known QSO (identified by \citet{Clark2005}) and X107 is clearly a background galaxy in the optical image, while the remaining X-ray sources have point-like optical counterparts, suggesting that they may be QSOs. A comparison between the X-ray and optical coordinates for these 10 sources gives a mean shift of $2\farcs5175$, and a standard deviation (i.e. $1\sigma$ positional uncertainties) of  $\approx0\farcs54$. This is similar to, although somewhat smaller than, the $\approx0\farcs56$ positional uncertainty between the X-ray positions and the WIRC IR images used in the \citet{Clark2011} study.

\section{Optical Star Cluster Counterparts to XRBs}

%%%%%%%%%%%%%%%%%%%%
\begin{deluxetable*}{cccccccl}
%\rotate
\tablecolumns{8}
\tablecaption{Optical Counterparts to X-Ray Sources\label{tbl-2}}
\tablewidth{0pt}
\tablehead{
\colhead{ID} & \colhead{Our Ages} & \colhead{Clark11+ Ages} & \colhead{Our Masses} & \colhead{Clark11+ Masses} & \colhead{$L_{X}$\tablenotemark{a}} & \colhead{S/N\tablenotemark{a}} & \colhead{Comments} \\
\colhead{} & \colhead{log($\tau$/yr)} & \colhead{log($\tau$/yr)} & \colhead{log($M/M_{\sun}$)} & \colhead{log($M/M_{\sun}$)} & \colhead{log(erg s$^{-1}$)} & \colhead{}\vspace{0.025 in} \\
\cline{1-8}\vspace{-0.06 in} \\
\multicolumn{8}{c}{High probability XRB candidates}
}
\startdata
%\cutinhead{High probability XRB candidates}
X27 & 6.7 & 7.00$-$7.48 & 5.2 & 5.05$-$5.58 & 39.77 & 644.1 & Weak H$_{\alpha}$ emission \\
X38 & 7.6 & 7.04$-$7.44 & 5.6 & 5.91$-$6.33 & 38.23 & 13.6 & 2$\sigma$ \\
X41 & 6.0 & \nodata & 5.5 & \nodata & 37.80 & 6.3 & Strong H$_{\alpha}$ emission \\
X45 & 7.7 & \nodata & 5.0 & \nodata & 37.40 & 3.8 & 2$\sigma$ \\
X47 & 6.7 & \nodata & 4.5 & \nodata & 38.35 & 37.0 & 2$\sigma$, weak H$_{\alpha}$ emission \\
X49 & 7.7 & \nodata & 4.9 & \nodata & 38.15 & 20.4 & Weak H$_{\alpha}$ emission nearby \\
X50 & 6.7 & \nodata & 4.9 & \nodata & 38.67 & 48.0  & 2$\sigma$, near NGC 4038 nucleus \\
X52 & 7.5 & \nodata & 5.8 & \nodata & \nodata & 1.6 & \\
X53 & 6.7 & \nodata & 5.0 & \nodata & 38.23 & 32.7 & Weak H$_{\alpha}$ emission \\
\nodata & 6.5 & \nodata & 4.9 & \nodata & \nodata & \nodata & \\
\nodata & 6.2 & \nodata & 4.6 & \nodata & \nodata & \nodata & \\
X58 & 6.8 & \nodata & 5.1 & \nodata & 38.41 & 25.7 & Weak H$_{\alpha}$ emission \\
X83 & 8.1 & 6.74$-$8.31 & 5.0 & 5.88$-$7.22 & 38.60 & 34.1 & 2$\sigma$ \\
X84 & 6.5 & \nodata & 5.3 & \nodata & 38.77 & 52.0 & H$_{\alpha}$ emission \\
\nodata & 6.6 & \nodata & 5.0 & \nodata & \nodata & \nodata & H$_{\alpha}$ emission \\
\nodata & 6.4 & \nodata & 4.8 & \nodata & \nodata & \nodata & H$_{\alpha}$ emission \\
X89 & 6.4 & 7.00$-$7.48 & 4.6 & 5.13$-$5.66 & 37.43 & 4.8 & Strong H$_{\alpha}$ emission \\
\nodata & 6.5 & \nodata & 5.4 & \nodata & \nodata & \nodata & 2$\sigma$ \\
X91 & 8.3 & \nodata & 6.2 & \nodata & 37.37 & 4.3 & \\
X98 & 7.6 & \nodata & 5.6 & \nodata & 37.60 & 4.5 & 2$\sigma$ \\
\nodata & 7.7 & \nodata & 5.1 & \nodata & \nodata & \nodata & 2$\sigma$ \\
\nodata & 8.1 & \nodata & 5.1 & \nodata & \nodata & \nodata & 2$\sigma$ \\
X99 & 8.2 & \nodata & 5.3 & \nodata & 39.54 & 254.4 & \\
\nodata & 6.7 & \nodata & 4.7 & \nodata & \nodata & \nodata & 2$\sigma$ \\
X101 & 6.6 & 7.12$-$7.19 & 5.7 & 5.97$-$5.99 & 38.02 & 10.3 & H$_{\alpha}$ emission \\
\nodata & 6.0 & \nodata & 4.9 & \nodata & \nodata & \nodata & H$_{\alpha}$ emission \\
X121 & 6.7 & \nodata & 5.5 & \nodata & 37.53 & 5.5 & H$_{\alpha}$ emission \\
\nodata & 6.5 & \nodata & 5.2 & \nodata & \nodata & \nodata & H$_{\alpha}$ emission \\
\cutinhead{Less certain XRB candidates}
X22 & 6.6 & \nodata & 4.7 & \nodata & 37.49 & 4.0 & 2$\sigma$, diffuse H$_{\alpha}$ emission \\
X24 & 6.5 & \nodata & 4.8 & \nodata & \nodata & 2.8 & Strong H$_{\alpha}$ emission \\
X36 & 6.6 & \nodata & 4.5 & \nodata & 37.97 & 9.0 & \\
X87 & 6.7 & 6.94$-$7.42 & 4.6 & 5.63$-$6.17 & 37.83 & 8.5 & Weak H$_{\alpha}$ emission
\enddata
\tablecomments{In the comments we distinguish clusters that are within 2$\sigma$ (0\farcs58) rather than 1$\sigma$ (0\farcs54) of the candidate XRB.}
\tablenotetext{a}{From Zezas et al. 2006.}
\end{deluxetable*}
%%%%%%%%%%%%%%%%%%%%

A comparison between the X-ray source positions and the optically selected star cluster catalog in the main part of the galaxy reveals that 22 XRBs (see Figure 2) are coincident with star clusters (15 XRBs have at least one cluster within $1\sigma$ or $0\farcs54$ and 7 others within  $2\sigma$ or $0\farcs58$). Therefore, in $\approx27\%$ of the XRB population we can uniquely identify or strongly constrain the parent star cluster. A $2\farcs55\times2\farcs55$ portion of the optical $HST$ image showing the location of these 22 X-ray sources and coincident clusters is shown in Figure 3. Their locations are also shown in Figure 2. We compile the properties of all coincident objects in Table 1. This includes the coordinates, signal-to-noise ratio (S/N), and luminosity for the XRBs from \citet{Zezas2006}, and the  cluster ages and masses estimated by \citet{Whitmore2010}. In a few cases, particularly in the crowded sites of recent star formation, we find more than a single potential cluster counterpart, and include information for all of these clusters, although in most cases they have similar ages. In cases where an XRB has more than one potential counterpart, we assume that the most massive cluster is the most likely host. In these cases we only list clusters located within 1$\sigma$ of the XRB.

Here, we check how many of the 22 candidate HMXBs found within $0\farcs58$ of a star cluster may be due to chance superposition. We populate the merger with 82 randomly distributed, synthetic XRBs in a fashion that follows the mass of the merger. This is accomplished by smoothing the F814W ($\approx I$ band) image, which best traces the overall mass of the Antennae and is less sensitive to recent star formation than shorter wavelength images, and using this as a weight map for the random simulations. These simulations were run 1000 times, and show $\approx 2-3$ chance coincidences between the location of XRBs and star clusters on average. This result suggests that the number of chance coincidences is significantly lower than the number that we observe, and hence almost all of the HMXBs that are coincident with star clusters are real associations and not due to chance superposition.

The remaining 60 out of 82 ($\approx73$\%) of candidate XRBs in the Antennae are not coincident with a star cluster. We have visually confirmed that none of these are coincident with clusters fainter than our adopted magnitude limit. In 5 of the remaining 60 cases, we do find a relatively bright point source within $1\sigma$ with $M_V$ between $\approx-7$ and $-9$, in the $HST$ images. These are likely the donor stars in the XRB. In the rest of the cases, no obvious optical counterpart to the candidate XRBs is observed.

\citet{Clark2011} found IR counterparts (within $2\arcsec$) to 32 of the X-ray sources listed in the \citet{Zezas2006} catalog that are within the main body of the Antennae (see their Figure 1). Excluding sources that are beyond the field of view of the ACS images used here, close to the northern nucleus, and  those that may be detections of diffuse plasma, we have 17 sources in common with \citet{Clark2011}.

Sources X45 and X49 are in our list but not in \citet{Clark2011}. They are coincident with clusters that have ages of $\approx50$~Myr and are somewhat fainter than most of the very young clusters in our sample, and hence may not have been detected in the WIRC observations used by \citet{Clark2011}. Meanwhile, \citet{Clark2011} list X7, X11, X86, and X94 as coincident sources, while we do not. These are missing from our coincident list for various reasons. For X11, we find an optical counterpart, but this appears to be a star, likely the donor star, rather than a cluster. X86 appears to fall in a dust lane in the ACS observations, with no obvious clusters nearby, while X94 is close to, but not coincident with ($\approx 2\arcsec$ away), a region of recent star formation which shows some H$\alpha$ emission, but no obvious clusters. \citet{Clark2011} find an extremely faint IR counterpart to source X7 which we do not detect. Based on these comparisons, our deep, high resolution  optical observations appear to be at least as effective as ground-based infrared imaging for identifying the counterparts to X-ray sources, despite the relatively large dust content of the Antennae. These observations have the added advantage of higher resolution, and hence the ability to age date the clusters more accurately.

X-ray point sources which are spatially coincident with star clusters are almost certainly XRBs. We note here that the X-ray images of the Antennae do reveal a significant amount of diffuse X-ray emission coincident with some of the X-ray sources listed in the \citet{Zezas2006} catalog, and that in a few cases the listed sources have poor contrast with this diffuse emission. Following \citet{Zezas2006}, we assume that these are likely faint XRBs, although it is possible that a few of the most marginal cases may be spurious detections within this diffuse emission. In Table~1 we list separately, at the end of the table, four X-ray sources from the \citet{Zezas2006} work which have a somewhat lower probability of being XRBs because of their low S/N in the X-ray images. The S/N from \citet{Zezas2006} is provided for all sources in Table~1.

The star clusters that are coincident with XRBs in the Antennae can be divided into three broad ranges in age. Of the 22 coincident sources, 14 of the counterparts are \emph{very young} with estimated ages of 6~Myr or younger, five of the counterparts are \emph{young} with ages of $\approx 20-50$~Myr, and three have \emph{intermediate} ages with $\tau\approx 100-300$ Myr. We discuss these three broad age categories in more detail below. XRBs coincident with very young clusters are found in the crowded star-forming regions of the Antennae, and as noted in Table~1, nearly all of these clusters appear to have at least some associated H$\alpha$ emission, which is a good indication of their youth. All of the clusters that likely host an XRB and which are younger than 10 Myr have masses higher than $3\times10^4~M_{\sun}$ and up to several $\times 10^5~M_{\sun}$, with a median mass of $\approx10^5~M_{\sun}$. The median X-ray luminosity of the XRBs with very young optical counterparts is $\approx10^{38}$~erg~s$^{-1}$.

Five of the X-ray sources are hosted by clusters that are somewhat older, with estimated ages of $\approx20-50$~Myr. As expected, clusters with these ages tend to be more spread out  within the Antennae than the very young clusters. The average and median mass of coincident clusters in this age range is somewhat higher than for the youngest clusters, with $\approx3\times10^5 M_{\sun}$. The X-ray luminosities of XRBs with $20-50$~Myr old counterparts range from $3\times 10^{37}$ to $2\times 10^{38}$~erg s$^{-1}$, with a median luminosity of $1.4\times10^{38}$ erg s$^{-1}$, i.e., roughly the same as the $<10$~Myr clusters.

Three X-ray sources (X83, X91, and X99) are hosted by even older clusters, which are $\approx100-300$~Myr old. Clusters with these ages are fairly well dispersed throughout the Antennae, and no longer trace regions of the most recent star formation. The clusters in this age range that are coincident with XRBs have estimated masses of $10^5$, $1.6\times 10^6$, and $2\times 10^5 M_{\sun}$, mostly higher than those of younger clusters that host XRBs. In fact, as we will discuss more fully in Paper~II, one-quarter of the (16) star clusters with ages of $100-300$~Myr and masses higher than $\approx 6 \times10^5~M_{\odot}$ are either coincident with or very close to an XRB, suggesting that clusters with these ages and masses are quite efficient at producing bright XRBs. The X-ray luminosities of the XRBs coincident with intermediate age clusters tend to be higher than those hosted by younger clusters, with luminosities of $4\times 10^{38}$, $2\times10^{37}$, and $3.5\times 10^{39}$~erg~s$^{-1}$, although the statistics are obviously poor.

\citet{Clark2011} presented ages for clusters coincident with 10 X-ray sources in the Antennae, one (X102) of which is outside of our field of view. Their age estimates are also listed in Table~1, and have fairly similar estimated ranges, log~$\tau \approx 6.9-7.5$ for nearly all of these clusters. We note that they do not derive ages as young as $3-6$~Myr for any cluster coincident with an XRB, as we have found here for the majority of coincident clusters.  Below, we compare the age results of coincident star clusters from \citet{Clark2011} with those from our work.

\begin{enumerate}

\item We find ages between 3 and 6~Myr for clusters associated with X87, X89, and X101. H$\alpha$ emission is observed in all three cases, as expected for clusters with these ages. \citet{Clark2011} find typical ages of $\tau \approx20 \pm 10$~Myr for clusters coincident with these sources.

\item We find an age between 20 and 50~Myr for the cluster associated with X38. \citet{Clark2011} find a similar age $\tau \thickapprox 20 \pm10$~Myr.

\item We find an age of $100-300$~Myr for the cluster associated with X83. \citet{Clark2011} find a (poorly constrained) age for this cluster somewhere between 5~Myr and 200~Myr.

\end{enumerate}

\noindent In addition to differences in the ages of coincident clusters, we derive systematically lower cluster masses than \citet{Clark2011}. The main reason for this discrepancy is likely the poorer resolution of the IR images compared with that of $HST$, $1\farcs50$ versus $\approx0\farcs51$, and the relatively large aperture radius used by \citet{Clark2011} for their photometry. It is clear from their Figure~2 and our Figure~1 that where they identify a single cluster, we sometimes find several, as they also note when comparing their IR images with the higher resolution optical images taken with WFPC2.

\section{Discussion}

\subsection{Constraining the Nature of XRBs within Star Clusters}

At an age of 6~Myr, stellar evolution models predict  that only stars initially more massive than $\approx 30~M_{\odot}$ will have evolved off the main sequence and become compact remnants. While there is still some uncertainty about the exact range of stellar masses that end their lives as NSs and those that become BHs, most models predict that, for lower metallicities, the transition between NSs and BHs occurs for stars with initial masses somewhere in the range of $\approx 18-25~M_{\odot}$ (e.g., \citet{Fryer1999}, \citet{Heger2003}), well below $30~M_{\odot}$. For approximately solar metallicities, as in the Antennae, stellar evolution models predict that massive stars develop substantial winds, which cause enough mass loss that the end product is a NS rather than a BH. Evolution of massive stars in tight binary systems however, is likely to be even more complicated with mass transfer back and forth, making it difficult for stellar evolution models
alone to predict which stars become BHs and which ones become NSs.

Independent clues to the nature of the compact objects in HMXBs come from their dynamics. $N$-body simulations show that NSs are far more susceptible to ejection from their parent cluster than BHs, due to natal kicks during the supernova explosion. \citet{PortegiesZwart2007} found that the vast majority of NSs ($\ge90\%$) are ejected from their parent clusters, while only $\approx45\%$ of the BHs were expelled.

We have recently performed $N$-body simulations of star clusters, with no primordial binaries, over their first $\approx300$~Myr of evolution, tracking all NSs and BHs (N. Sen et al. 2012, in preparation). Our results confirm those of \citet{PortegiesZwart2007}, and extend them to candidate HMXBs, which form dynamically in these simulations. Our main result is that an HMXB \emph{found within its parent star cluster} is very likely to have a BH and not a NS as the compact object (this is true for ages up to at least $\approx300$~Myr, the limit of our simulations). The reason for this is twofold: (1) BHs have a much higher probability of being retained within their parent clusters (as also found by \citet{PortegiesZwart2007}); and (2) the higher masses of the retained BHs, when compared with NSs, make them much more likely to form a binary system (e.g., \citet{Garofali2012}, \citet{Converse2011}). Conversely, HMXBs found in the field and not with their parent clusters or associations are much more likely to have a NS as the compact object. It is possible that a population of primordial binaries may somewhat alter these conclusions, but this will depend on the initial mass ratios and orbital parameters of the binaries. In a separate work, we will explore how these parameters of primordial binaries affect the relative fraction of BH versus NS X-ray-emitting binaries, both in and out of clusters.

\subsection{The Nature of HMXBs Associated with Star Clusters in the Antennae}

In Section~5.1 we presented dynamical arguments that suggest that HMXBs hosted by star clusters have a BH as the compact object, regardless of the metallicity-dependent evolutionary path that led to the formation of the compact object. We therefore conclude that most of the candidate HMXBs studied in this work, i.e., those coincident with star clusters, have a BH as the compact object, because they still reside within their parent clusters.  This suggests that at least a quarter (22 out of 82) of the luminous XRBs observed in the Antennae have a BH as the compact object. Although quite uncertain, most XRBs in the Galaxy and the Magellanic Clouds appear to have a NS as the compact object. According to \citet{Belczynski_Ziolkowski2009} NS XRBs outnumber BH XRBs by a factor of $\sim30$ in our Galaxy. These X-ray-emitting binaries are found primarily in the field, and not in clusters/associations, and hence the large fraction of NS-binaries is consistent with the dynamical arguments presented above.

The ages of $\approx 3-6$~Myr determined for very young star clusters in the Antennae that are coincident with HMXBs is consistent with predictions of HMXB formation. \citet{Linden2010} followed the evolution of primordial pairs of stars drawn from a Salpeter ($\alpha=-2.35$) initial mass function, with a minimum mass of $4~M_{\odot}$ for the primary, and the mass for the secondary drawn randomly from a flat mass ratio distribution for the mass ratios. The initial binary separation was drawn from a flat distribution (in log space) with an upper limit of $10^{5}R_{\odot}$. They use the population synthesis code \emph{StarTrack} \citep{Belczynski2008} to model processes such as stable mass transfer through Roche lobe overflow and unstable common envelope phases. \citet{Linden2010} show in their Figure~1 that in solar metallicity systems like the Antennae, bright ($L_X > 10^{36}~\mbox{erg~s}^{-1}$) HMXBs turn on suddenly at ages of approximately 4~Myr, with a production rate that is sharply peaked between $4$ and $6$~Myr and dropping off thereafter. These ages are quite similar to those listed in Table~1 for many coincident clusters, and are probably not due to a peak in the star formation rate $4-6$~Myr ago, since: (1) dynamical simulations which reproduce the observed morphology of the merging Antennae do not find such a peak in the star formation rate (see e.g., \citet{Karl2011}), and (2) it would be impossible to have such a well timed burst over the entire field of the Antennae since the communication time is $\sim10^{9}$ yr for a signal traveling at the typical random velocity of the interstellar medium of $\approx10$ km s$^{-1}$ (e.g., \citet{Fall2009,Whitmore2007}). Hence, our age analysis is consistent with predictions for the production time scale of HMXBs in high metallicity galaxies.

In the dwarf starburst galaxy NGC~4449 \citep{Rangelov2011} we found three HMXBs coincident with very young star clusters and one coincident with an intermediate-age one. Although NGC 4449 has a smaller number of HMXBs than the Antennae (since it has fewer stars in general) leading to poorer statistics when analyzing different HMXB subpopulations, there do appear to be similarities between the two systems. In both cases, $\approx15\%$ of all HMXBs are coincident with very young clusters ($\tau \lea 8$ Myr). There are however, no HMXBs in NGC~4449 coincident with $30-50$ Myr old clusters. We will further investigate this issue by adding more dwarf starburst galaxies to our sample in the future.

The Antennae galaxies contain 6 ultra-luminous X-ray sources (ULXs, $L_{X}>10^{39}$~erg~s$^{-1}$). Of these we find 2 that are coincident with star clusters: X27 with a very young ($\tau\approx3-6$~Myr) cluster and X99 with an intermediate-age ($\tau\approx100-200$~Myr) cluster. 

\section{Summary and Conclusions}
 
In this work, we compared the locations of 82 candidate XRBs (from \citet{Zezas2006}) with the locations of massive star clusters (from \citet{Whitmore2010}) in the merging Antennae galaxies. Our main conclusions are as follows.

\begin{enumerate}

\item 22 out of the 82 candidate XRBs are located within $0\farcs58$ of a star cluster, i.e. within the $2\sigma$ positional uncertainty of the X-ray and optical observations. Only 2-3 coincidences between XRBs and star clusters would be expected due to chance superposition, indicating that the XRBs likely formed within these star clusters.

\item We found that the ages of host clusters fall within three ranges: (1) very young, with $\tau\approx3-6$~Myr; (2) young, with $\tau\approx30-50$~Myr; and (3) intermediate age, with $\tau\approx100-300$~Myr. Fourteen of the 22 XRBs are hosted by very young clusters, five by young clusters, and three by intermediate-age clusters. 

\item We found two ULX candidates that are coincident with a star cluster (one cluster has an age of $\tau\approx5-6$~Myr and the other $\tau\approx100-200$~Myr).

\item Direct $N$-body simulations of star clusters have shown that XRBs found within their parent clusters are highly likely to have a BH rather than a NS as the compact object. We therefore conclude that the 22 XRBs that are coincident with a cluster in the Antennae are most likely BH binaries of different ages.

\end{enumerate}

Our results indicate that the relative locations of XRBs and star clusters combined with precise age dating of star clusters are a powerful approach for finding BH binaries and determining their ages.

\acknowledgements

We thank the referee, Dr. Stephen Eikenberry, whose careful reading and helpful suggestions significantly improved our manuscript. We also thank Dr. Joseph Converse for helpful discussions on star cluster dynamics. This work was supported by NASA contract NAS8-39073 (CXC) and NASA AR8-9010.

%%%%%%%%%%%%%%%%%%%%

{}

\end{document}